\documentclass[10pt,journal,twocolumn]{IEEEtran}
\IEEEoverridecommandlockouts
\usepackage{cite}
\usepackage{amsmath,amssymb,amsfonts,booktabs}
\usepackage{algorithmic}
\usepackage{graphicx}
\usepackage{textcomp}
\usepackage{xcolor}
\usepackage{enumerate}
\usepackage{epstopdf}
\usepackage{subcaption}
\usepackage{adjustbox}

\usepackage{amsfonts}

\usepackage{amsthm}
\usepackage{multirow}
\usepackage{makecell}
\usepackage{color}
\usepackage{bm}

\usepackage{mathtools}

\usepackage{float}
\usepackage{setspace}

\allowdisplaybreaks[4]

\newtheorem{proposition}{Proposition}

\usepackage{algorithm}
\usepackage{algorithmic}

\begin{document}

\title{\LARGE UAV-Enabled Data Collection for IoT Networks via Rainbow Learning}

\author{
        Yingchao Jiao,
        Xuhui Zhang,
        Wenchao Liu,
        Yinyu Wu,
        Jinke Ren,
        Yanyan Shen, 
        Bo Yang,
        and Xinping Guan


\thanks{
Y. Jiao and Y. Wu are with the Shenzhen Institute of Advanced Technology (SIAT), Chinese Academy of Sciences, Guangdong 518055, China, and are also with the University of Chinese Academy of Sciences, Beijing 100049, China (e-mail: yc.jiao@siat.ac.cn; yg.wu@siat.ac.cn).
}

\thanks{
X. Zhang and J. Ren are with the Shenzhen Future Network of Intelligence Institute, the School of Science and Engineering, and the Guangdong Provincial Key Laboratory of Future Networks of Intelligence, The Chinese University of Hong Kong, Shenzhen, Guangdong 518172, China (e-mail: xu.hui.zhang@foxmail.com; jinkeren@cuhk.edu.cn).
}


\thanks{
W. Liu and Y. Shen are with the SIAT, Chinese Academy of Sciences, Guangdong 518055, China (e-mail: wc.liu@foxmail.com; yy.shen@siat.ac.cn).
}

\thanks{B. Yang and X. Guan are with the Department of Automation and the Key Laboratory of System Control and Information Processing, Ministry of Education, Shanghai Jiao Tong University, Shanghai 200240, China (e-mail: bo.yang@sjtu.edu.cn; xpguan@sjtu.edu.cn).}

\vspace{-3.5em}
}

\maketitle
	
\begin{abstract}
Unmanned aerial vehicles (UAVs) enabled Internet of things (IoT) systems have become an important part of future wireless communications. To achieve higher communication rate, the joint design of UAV trajectory and resource allocation is crucial.
In this paper, a multi-antenna UAV is dispatched to simultaneously collect data from multiple ground IoT nodes (GNs) within a time interval.
To improve the sum data collection (SDC) volume from the GNs, the UAV trajectory, the UAV receive beamforming, the scheduling of the GNs, and the transmit power of the GNs are jointly optimized.
Since the problem is non-convex and the variables are highly coupled, it is hard to be solved using traditional methods.
To find a near-optimal solution, a double-loop structured optimization-driven deep reinforcement learning (DRL) algorithm, called rainbow learning based algorithm (RLA), and a fully DRL-based algorithm are proposed to solve the problem effectively.
{Specifically, the outer-loop of the RLA utilizes a fusion deep Q-network to optimize the UAV trajectory, GN scheduling, and power allocation, while the inner-loop optimizes receive beamforming by successive convex approximation.}
Simulation results verify that the proposed algorithms outperform two benchmarks with significant improvement in SDC volumes, {energy efficiency, and fairness}.
\end{abstract}
	
\section{Introduction}
With the rapid advancement of the Internet of things (IoT) technology, an increasing number of IoT devices and sensors have been deployed in wireless networks, generating and collecting a large amount of data \cite{9456851}. The data contain rich information that can be utilized for monitoring, managing, and providing services, such as urban traffic, environmental monitoring, and agricultural management \cite{9779853}. However, traditional IoT technology has limited data collection capability due to the limited coverage and the difficulty of transmitting in complex terrains. To address this issue, unmanned aerial vehicle (UAV) technology offers a new solution. UAVs, with flexible maneuverability and extensive coverage, can freely fly in the air and provide line-of-sight (LoS) links for data collection and transmission. Consequently, integrating UAVs with IoT systems to achieve the data collection mission has become a key point of current research \cite{9273074}.

Several pioneering efforts have been dedicated to addressing the challenge of the UAV data collection in IoT applications \cite{9712630, shen2023average, 10233692, 10606316, 10568916}.
In \cite{9712630}, the authors proposed a combination framework of transfer learning and deep reinforcement learning (DRL) to jointly optimize the flying speed and energy replenishment of the UAVs.
To minimize the age of information (AoI) for all IoT devices, the authors in \cite{shen2023average} designed a joint optimal data collection time, transmit power and trajectory strategy for the UAV.
In \cite{10233692}, the completion time was minimized in a single-antenna UAV data collection system, where the collection order and the transmit power of the ground IoT nodes (GNs) were optimized.
The single-antenna UAV system was studied in \cite{10606316}, where the minimum computation data was optimized by jointly designing the UAV trajectory, and the resource allocation.
Besides, the UAV’s data collection time and energy consumption were minimized by jointly optimizing the UAV’s collection point, flight trajectory, and flight speed \cite{10568916}.
However, existing works face some limitations. Firstly, each UAV is equipped with a single antenna, limiting the system capacity for achieving higher data collection rate, and leading to a lower quality of service.
Second, based on the assumption that the hovering position of the UAV for data collection is pre-determined, the UAV trajectory optimization is typically modeled as a traveling salesman problem, and the sequence of visits to each hovering location needs to be determined.
Under this assumption, the maneuverability and flexibility of the UAVs cannot be fully utilized.

To overcome the above two issues, this paper explores a data collection system where a UAV equipped with multiple antennas is dispatched to collect data from GNs. To enhance the achievable sum data collection (SDC) volume, we jointly optimize the UAV trajectory, the receive beamforming, the GN scheduling, and the transmit power of all GNs.
Firstly, we introduce a double-loop structured optimization-driven DRL algorithm called rainbow learning algorithm (RLA), where the UAV receive beamforming is optimized in the inner-loop, and other variables are learned in the outer-loop.
{Leveraging this double-loop architecture, the proposed RLA synergistically combines the strengths of successive convex approximation (SCA) in the inner-loop and the DRL in the outer-loop, achieving significant performance gains in communication systems.}
Second, we propose a fully DRL-based approach, called all learned once, to optimize the achievable data
collection volume of all GNs.
Numerical results verify that the proposed two algorithms achieve significant performance improvements compared with two baseline algorithms.

\section{System Model and Problem Formulation}
\subsection{Network and UAV movement model}
\begin{figure}[htbp]
    \centering 
    \includegraphics[width=0.4\textwidth]{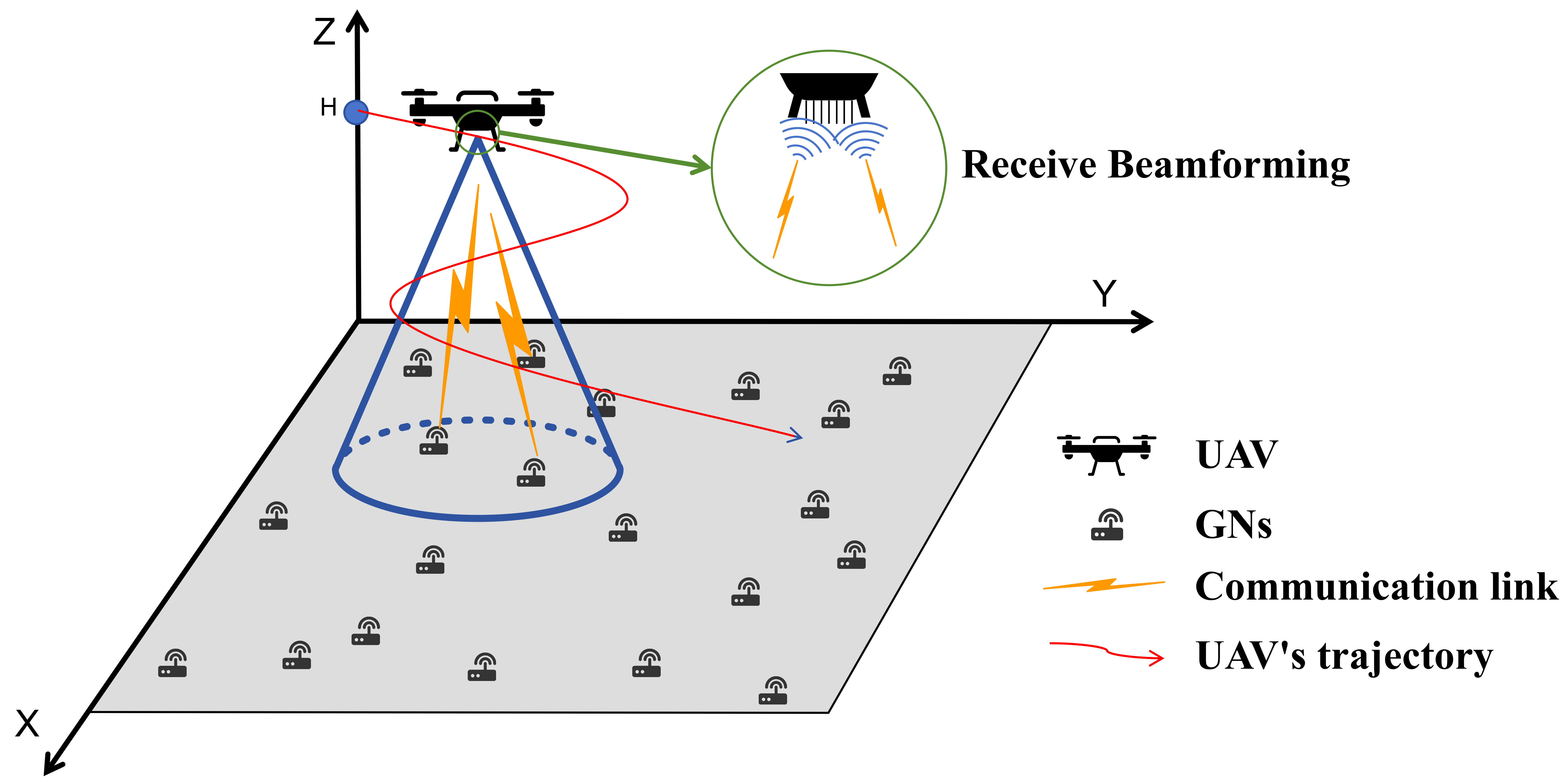} 
    \caption{System model of the UAV-enabled IoT data collection.} 
    \label{fig:system_model} 
\end{figure}

{As shown in Fig.~\ref{fig:system_model}}, we consider an IoT data collection system, where one UAV flies over $K$ GNs to collect data from the GNs within $T$ seconds. The set of the GNs is denoted as $\mathcal{K} = \{1,2,\cdots, K\}$. The UAV is equipped with $M$ antennas. Each GN is equipped with a single antenna. We assume that the locations of all GNs are fixed. The location of the $k$-th GN is given by $(\boldsymbol{s}_k, 0 )$, where $\boldsymbol{s}_k = \{ s_k^{\mathrm{x}}, s_k^{\mathrm{y}} \}$ denotes the two-dimensional (2D) coordinates of the $k$-th GN. The total time interval is equally divided into $N$ time slots, each of which has a duration of $\tau = \frac{T}{N}$ seconds, and the set of the time slots is denoted as $\mathcal{N} = \{1,2,\cdots, N\}$. The position of the UAV at time slot $n$ is given by $(\boldsymbol{q}[n], \mathcal{H} )$, where $\boldsymbol{q}[n] = ( q^{\mathrm{x}} [n], q^{\mathrm{y}} [n])$ is the 2D coordinates of the UAV, and $\mathcal{H}$ is the fixed flying altitude.

The current velocity of the UAV at time slot $t$ is given by $\boldsymbol{v}[n] = \frac{1}{\tau}\left (\boldsymbol{q}[n]-\boldsymbol{q}[n-1]\right ), \ 2\leq{n}\leq{N}$,
where the velocity of the UAV at the first time slot is initialized by a pre-defined value.
To ensure that the UAV can collect data from GNs within the time interval, the UAV should comply with the minimum velocity $V_{\min}$, maximum velocity $V_{\max}$, and maximum acceleration $a_{\max}$,
which are given by
\begin{equation} 
{V_{\min}}\leq \left \Vert \boldsymbol{v}[n] \right \Vert  \leq{V_{\max}}, \quad2\leq{n}\leq{N}. 
\label{v1}  
\end{equation}  
\begin{equation}   
{0}\leq \left \Vert\frac{\boldsymbol{v}[n]-\boldsymbol{v}[n-1]}{\tau}\right \Vert\leq{a_{\max}}, \quad3\leq{n}\leq{N}.  
\label{v2}  
\end{equation}  

\subsection{Channel model}
The channel matrix from the GNs to the UAV is $\boldsymbol{{H}}_{\mathrm{UAV}}[n] = \{\boldsymbol{h}_1[n],\cdots,\boldsymbol{h}_k[n],\cdots,\boldsymbol{h}_K[n] \} \in \mathbb{C}^{M\times K}$, where $\boldsymbol{h}_k[n]$ is the channel corresponding to the $k$-th GN at time slot $n$, which is expressed as
\begin{equation} 
\boldsymbol{h}_k[n]=\sqrt{\frac{\chi_k[n]}{{C}}}\boldsymbol{\Theta}_k[n] \boldsymbol{\xi}_k[n],
\end{equation}
where $\chi_k[n]$ denotes the large-scale fading coefficient, ${C}$ represents the number of multiple path components, where the set of the paths are denoted by $\mathcal{C} = \{1, \cdots, C\}$, $\boldsymbol{\Theta}_k[n] = \{ \boldsymbol{\theta}_k^1 [n],\cdots,\boldsymbol{\theta}_k^c [n] ,\cdots, \boldsymbol{\theta}_k^C [n]  \}$ is the array response matrix. The $c$-th path steering vector of the $k$-th GN is given by $\boldsymbol{\theta}_k^c [n] = \{ 1, \cdots, e^{\mathrm{j}\pi \sin(\varrho_k^c [n])},\cdots, e^{\mathrm{j}\pi (M-1)\sin(\varrho_k^c [n])} \}^{\mathsf{T}}$, where $\mathrm{j}$ is the imaginary unit, $\varrho_k^c [n]$ denotes the angle of arrival (AoA) from the $k$-th GN to the UAV on the $c$-th path. $\boldsymbol{\xi}_k[n] = \{ \xi_k^1[n],\cdots,\xi_k^c[n],\cdots, \xi_k^C [n] \}^{\mathsf{T}}$ denotes the small-scale fading coefficients of all paths.

\subsection{Data collection model}
The coverage state of the UAV
is introduced by a binary indicator $\boldsymbol{1}_k [n]$,
where the value of $\boldsymbol{1}_k [n]$ is set to $1$, if the distance from the $k$-th GN to the UAV is smaller than a pre-defined threshold $d_{\mathrm{th}}$. Otherwise, it is set to 0.
Hence, the set of the scheduled GNs by the UAV at time slot $n$ is given by $\mathcal{K}[n] = \{ k \vert \boldsymbol{1}_k [n] = 1, \forall k\} $. In the following, we denote the the $k$-th GN in the coverage of the UAV, i.e., $\boldsymbol{1}_k [n]$ = $1$, in the set $ \mathcal{K}[n]$, for simplicity. 
Thus, the received signal $y[n]$ is given by
\begin{equation} 
y[n] = \boldsymbol{w}_k^{\mathsf{H}}[n] \left ( \sum_{k\in \mathcal{K}[n]} \sqrt{P_k[n]}  \boldsymbol{h}_k[n]s_k[n] + \Tilde{\boldsymbol{n}}[n] \right ), 
\end{equation}
where $P_k[n]$ denotes the transmit power of the $k$-th GN, bounded by power capacity $P_{k}^{\max}$, i.e., $P_k[n] \le P_{k}^{\max}$, 
$\boldsymbol{w}_k[n] \in \mathbb{C}^{M\times 1}$
is the receive beamforming vector associated with the $k$-th GN at the UAV, $s_k[n]$ is the data signal of the $k$-th GN with unit power, i.e., $\mathbb{E}\{\vert s_k[n]
\vert^{2}\} = 1$, and $\Tilde{\boldsymbol{n}}[n] = \{ \Tilde{n}_1 [n],\cdots, \Tilde{n}_m [n],\cdots, \Tilde{n}_M [n] \}^{\mathsf{T}}$ denotes the noise vector.
Meanwhile, we have ${\left\Vert\boldsymbol{w}_k[n]\right\Vert \le 1,\quad \forall n\in \mathcal{N}}\,\label{p1f}$.

Hence, the signal-to-interference-plus-noise ratio (SINR) of the $k$-th GN's signal can be expressed as
\begin{equation} 
\gamma_k [n] (\boldsymbol{w}_k, \boldsymbol{q}, P_k) = \frac{P_k[n]{{\vert}{\boldsymbol{w}_k^{\mathsf{H}}[n] \boldsymbol{h}_k[n]}{\vert}}^2}{\mathcal{P}_k [n]+{\sigma}^2\Vert \boldsymbol{w}_k[n]\Vert^2},
\end{equation}
where $\mathcal{P}_k [n] = \sum_{j \in \{\mathcal{K}[n] \backslash k\}}{{P_j[n]{{\vert}{\boldsymbol{w}_k^{\mathsf{H}}[n] \boldsymbol{h}_j[n]}{\vert}}^2}}$ is the interference from other GNs, and $\sigma^2$ is the noise power. Therefore, the data collection rate of the $k$-th GN can be expressed as
\begin{equation} 
R_k [n] (\boldsymbol{w}_k, \boldsymbol{q}, P_k) =  \log_2 (1 + \gamma_k [n] (\boldsymbol{w}_k, \boldsymbol{q}, P_k)).
\end{equation}
Accordingly, the achievable data volume of the $k$-th GN can be written as
\begin{equation} {
    D_k[n](\boldsymbol{w}_k, \boldsymbol{q}, P_k) =
    \begin{cases}
    \tau B R_k[n](\boldsymbol{w}_k, \boldsymbol{q}, P_k),  &\text{if}\ \boldsymbol{1}_k [n]=1,\\
    0, &\text{otherwise},
    \end{cases}}
\end{equation}
where $B$ is the system bandwidth. Besides, to ensure fairness of all GNs to access the UAV, the minimum data collection volume for GNs should be larger than a pre-defined threshold $D_{\mathrm{th}}$. Thus the fairness constraint is defined by
\begin{equation} 
\sum_{n=1}^N D_k[n](\boldsymbol{w}_k, \boldsymbol{q}, P_k) \geq D_{\mathrm{th}}.
\label{fairness_constraint}
\end{equation}

\subsection{Problem formulation}
The achievable SDC volume of the UAV during one time interval is defined by
\begin{equation} 
\mathcal{U}(\boldsymbol{W}, \boldsymbol{Q}, \boldsymbol{P}) = \sum_{n=1}^N \sum_{k=1}^K D_k[n](\boldsymbol{w}_k, \boldsymbol{q}, P_k),
\label{SDC}
\end{equation}
where \scalebox{0.94}{$\boldsymbol{W} \triangleq \{ \boldsymbol{w}_k [n], \forall k \in \mathcal{K}[n], \forall n\}$}, \scalebox{0.94}{$\boldsymbol{Q} \triangleq \{ \boldsymbol{q} [n], \forall n\}$}, and \scalebox{0.94}{$\boldsymbol{P} \triangleq \{ {P}_k [n], \forall k, \forall n\}$} are the decision variables for the receive beamforming at the UAV, the trajectory of the UAV, and the power allocation of the GNs, respectively.

We aim to maximize the SDC volume to improve the system capacity, by jointly optimizing the
decision variables including $\boldsymbol{W}$, $\boldsymbol{Q}$, and $\boldsymbol{P}$.
Therefore, the optimization problem can be formulated as {\begin{subequations}{\begin{flalign} 
(\text{P1}) \max_{\boldsymbol{W}, \boldsymbol{Q},\boldsymbol{P} } \quad & \mathcal{U}(\boldsymbol{W}, \boldsymbol{Q}, \boldsymbol{P})  \nonumber\\
 {\rm{s.t.}}  \quad 
&\eqref{v1},\eqref{v2},\eqref{fairness_constraint},\nonumber\\
& P_k[n] \le P_{k}^{\max},  \quad \forall k\in \mathcal{K},\quad \forall n\in \mathcal{N},\label{p1e}\\
& {\left\Vert\boldsymbol{w}_k[n]\right\Vert \le 1,\quad \forall k\in \mathcal{K},\quad \forall n\in \mathcal{N}},\label{p1f}
\end{flalign}}\end{subequations}}where constraints \eqref{v1} and \eqref{v2} are the UAV movement constraints, \eqref{fairness_constraint} denotes the data collection fairness of all GNs,
\eqref{p1e} is the power constraint of the GNs.
\eqref{p1f} is the feasible region constraint of the receive beamforming.
{Since the objective function of problem P1 has the coupling of multiple optimization variables, and the left hand side of the constraint condition (8) is a non-concave function, problem P1 is non-convex, and thus the problem is very difficult to solve  by traditional optimization methods in polynomial time. In the following, we will propose a double-loop structured algorithm that combines DRL and optimization methods. To facilitate reading, the main notations are summarized in Table \ref{tab:variables_description}.}

\begin{table}[!htbp] \scriptsize
\centering
\caption{{Main Notations.}}
\begin{tabular}{cc}
\toprule
\centering
Notation& Description\\
\midrule  %
\centering $\tau$ & Time slot duration\\
\centering $\boldsymbol{h}_k[n]$ & Channel gain of the $k$-th GN at time slot $n$\\
\centering $\boldsymbol{w}_k[n]$ & Receive beamforming associated with the $k$-th GN at the UAV\\
\centering $y[n]$ & Received signal at time slot $n$\\
\centering $\sigma^2$ & Noise power\\
\centering $B$ & System bandwidth\\
\centering $P_k[n]$ & {Transmit power of the $k$-th GN at time slot $n$}\\
\centering $\mathcal{P}_k[n]$ & Interference from other GNs at time slot $n$\\
\centering $P_k^{\max}$ & Maximum transmit power of the $k$-th GN\\
\bottomrule
\end{tabular}
\label{tab:variables_description}
\end{table}

\section{UAV-Enabled Data Collection Systems via Rainbow Learning}
To solve problem (P1), we introduce 
the RLA,
which is a double-loop structured algorithm composed of an enhanced fusion deep Q-learning network (FDQN) and the SCA method.
Specifically,
the trajectory of the UAV, and the scheduling and power allocation of the GNs are optimized at the outer-loop by the FDQN, while the receive beamforming of the UAV is optimized at the inner-loop by the SCA.
\subsection{Outer-loop Learning}
\paragraph{Reformulation of (P1)}
DRL algorithms aim to optimize a policy
that chooses
actions through the interaction of an agent with its environment to maximize an expected reward. The interaction is illustrated as a tuple of a Markov decision process (MDP), including the state $s$, which is the observation of the environment, the action $a$, which denotes the
determined optimization varibles according to the problem, and the reward $r$, which evaluates the action according to the state.

The outer-loop of RLA aims to optimize the trajectory of the UAV, the scheduling, and power allocation of the GNs.
Therefore, we reformulate problem (P1) as an MDP. The components of the MDP are given as follows.

\textit{Observation state space \( \mathcal{S} \)}:
The observation state space consists of the current position of the UAV, the channel state information, and the positions of all GNs.
Therefore, the observation state space can be represented as
\begin{equation} 
\mathcal{S}[n] = (\boldsymbol{q}[n], \boldsymbol{H}_{\mathrm{UAV}}[n], \boldsymbol{s}_{k}),\ \forall k,\ \forall n.
\end{equation}

\textit{Action space \( \mathcal{A} \)}:
The action space can be represented as
\begin{equation} 
 \mathcal{A}[n] = (\Delta_{\text{UAV}}[n], \theta_{\text{UAV}}[n],  \boldsymbol{P}_{\text{}}[n]),\ \forall n,
\end{equation}
    where $\Delta_{\text{UAV}}[n]$ denotes the moving distance of the UAV, and \( \theta_{\text{UAV}}[n] \) represents the flight direction angle of the UAV. With the decision of the movement of the UAV, the scheduling of the GNs can be further determined. In order to compress the dimensions of the solution space and improve exploration efficiency, all actions in the action space $\mathcal{A}$ are discretized.

\textit{Reward function \( R \)}: The objective of problem (P1) is to maximize the SDC volume collected by the UAV.
Hence, we define the reward function as
\begin{equation} 
 {r[n]} = \sum_{i=1}^n \sum_{k=1}^K D_k[i](\boldsymbol{w}_k, \boldsymbol{q}, P_k) +
        {R}_{\text{cover}} [n](\boldsymbol{1}_k) - {P}_{\text{out}},
\label{calculate_reward}
\end{equation}
where \( {R}_{\text{cover}} [n](\boldsymbol{1}_k) = \omega\sum_{k=1}^K\min(1,\sum_{i=1}^n \boldsymbol{1}_k)\) is the bonus reward w.r.t the number of GNs that have been scheduled by the UAV, $\omega$ is the weighted value.
\( {P}_{\text{out}} \) represents the penalty for the UAV moving outside the restricted area.

\paragraph{Fusion Deep Q-learning Network}
DRL algorithms are well-suited for addressing various complex problems like problem (P1) \cite{8714026}.
To overcome the over-estimation and stability issues in former DRL algorithms, FDQN is proposed, which is a fusion of the dueling network, the noisy network, the double network structure, and the prioritized experience replay (PER) \cite{hessel2018rainbow}, which are introduced respectively in the following.

\textit{Dueling Network}: the ${Q}$-value is decomposed into a state-value function \( V \) and an advantage function \( A \), i.e., \( Q = V + A \), where \( A \) measures the difference between the value of taking action \( a \) in state \( s \) and the value of the state \( s \) itself. This decomposition reduces the impact of action space on the ${Q}$-value and balances the state-value and action-value.

\textit{Noisy Network}: The exploration and robustness are improved by adding learnable Gaussian noise to the weights of the fully connected layers. This randomness promotes diverse exploration and enhances the network's ability to learn environmental features and patterns.

\textit{Double Network}:
To stabilize training and avoid over-estimation, two networks are utilized: the evaluation network $Q$ for action selection, and the target network $Q'$ for evaluation by calculating target value \scalebox{0.94}{$
 y' = r + \gamma \cdot Q'(s^-, \arg\max_a(Q(s^-, a)))
$},
where \( \gamma \) is the discount factor, and $s^-$ is the state of the next time slot. 

\textit{PER}: After each interaction, the MDP tuple is stored in the PER buffer. During training, a batch of tuples are sampled from the buffer to update the evaluation network. Its parameters are periodically soft-updated with update rate \scalebox{0.94}{$\Bar{\tau}$} to the target network as \scalebox{0.94}{$\theta' \leftarrow \Bar{\tau} \theta + (1 - \Bar{\tau}) \theta'$}, where \scalebox{0.94}{$\theta$} and \scalebox{0.94}{$\theta'$} are the evaluation and target network parameters, respectively.

\subsection{Inner-loop Optimization}
The inner-loop of RLA aims to optimize the receive beamforming of the UAV. Thus, problem (P1) is simplified as
{\begin{subequations} {
\begin{flalign}
(\text{P2})\ \max_{\boldsymbol{W}} \quad & \sum_{k\in \mathcal{K}[n]} R_k [n] (\boldsymbol{w}_k) \nonumber\\
 {\rm{s.t.}}  \quad
& \eqref{fairness_constraint}\ \text{and}\ \eqref{p1f}. \nonumber
\end{flalign}}\end{subequations}}Let {$\boldsymbol{W}_k[n] = \boldsymbol{w}_k[n]\boldsymbol{w}_k^{\mathsf{H}}[n]$}, where {$\boldsymbol{W}_k[n] \succeq \boldsymbol{0}$} and {$\mathrm{rank}(\boldsymbol{W}_k[n]) \leq 1$}.
Since the UAV trajectory and transmit power of the GNs are fixed, 
Eq. \eqref{SDC} can be expressed as the SDC per time slot.
Let {$\Bar{\mathcal{N}}_k \triangleq \{ \Bar{n} \vert \arg_{n} (\boldsymbol{1}_k[n] = 1) \}$} denote the time index set that the $k$-th GN is scheduled.
Thus, (\text{P2}) can be rewritten as
{\begin{subequations} {
\begin{flalign}
(\text{P3})\ \max_{\boldsymbol{W}} \quad & \sum_{k\in \mathcal{K}[n]} R_k [n] (\boldsymbol{W}_k) \nonumber\\
 {\rm{s.t.}}  \quad
&\mathrm{tr} \left(\boldsymbol{W}_k[n]\right) \le 1, \label{p3a}\\
&\mathrm{rank}(\boldsymbol{W}_k[n]) \leq 1,\quad \forall k\in \mathcal{K}[n], \label{p3b}\\
&\sum_{n\in \Bar{\mathcal{N}_k}} R_k[n](\boldsymbol{w}_k, \boldsymbol{q}, P_k) \geq \frac{D_{\mathrm{th}}}{\tau B }, \quad \forall k \in
\mathcal{K},\label{p3c}
\end{flalign}}\end{subequations}}where the objective function $R_k [n] (\boldsymbol{W}_k)$ is given by
\begin{equation}{\footnotesize
    \begin{split}
        R_k [n] (\boldsymbol{W}_k) = \log_2 \left (\mathrm{tr}\left(\sum_{i \in \mathcal{K}[n]}P_i[n]\boldsymbol{H}_k[n]\boldsymbol{W}_k[n] \right)
    + \mathrm{tr} (\sigma^2 \boldsymbol{W}_k[n]) \right)\\
    - \log_2 \left(
        \sum_{j \in \{\mathcal{K}[n] \backslash k\}}
    \mathrm{tr}(P_j[n]{\boldsymbol{H}_j[n]}\boldsymbol{W}_k[n] )
    + \mathrm{tr} (\sigma^2 \boldsymbol{W}_k[n])
    \right),
    \end{split}
    \label{concave}}
\end{equation}
where \scalebox{0.94}{$\boldsymbol{H}_k[n] = \boldsymbol{h}_k[n]\boldsymbol{h}_k^{\mathsf{H}}[n], 
\forall k \in \mathcal{K}[n]$}. The objective function of problem (P3) is non-convex and has a concave-minus-concave
form. We introduce Taylor expansion on the second concave term in \eqref{concave}, and then approximate \scalebox{0.94}{$R_k [n] (\boldsymbol{W}_k)$} by its lower bound \scalebox{0.94}{$\check{R}_k [n] (\boldsymbol{W}_k)$} in the $l$-th iteration of SCA, which is given by
\begin{equation}{\footnotesize
    \begin{split}
        &R_k [n] (\boldsymbol{W}_k) \geq \log_2 \left (\mathrm{tr}(\sum_{i \in \mathcal{K}[n]}P_i[n]\boldsymbol{H}_k[n]\boldsymbol{W}_k[n] )
    + \mathrm{tr} (\sigma^2 \boldsymbol{W}_k[n]) \right)\\
    &- \log_2 \left(
        \sum_{j \in \{\mathcal{K}[n] \backslash k\}}
    \mathrm{tr}(P_j[n]\boldsymbol{H}_j[n]\boldsymbol{W}_k^{(l)}[n] )
    + \mathrm{tr} (\sigma^2 \boldsymbol{W}_k^{(l)}[n])
    \right)\\
    &-\sum_{j \in \{\mathcal{K}[n] \backslash k\}}
    \mathrm{tr}
        \left ( \frac{
            \log_2(\mathrm{e}) \left (
            P_j[n]\boldsymbol{H}_j[n] \left ( \boldsymbol{W}_k[n]-\boldsymbol{W}_k^{(l)}[n] \right ) 
            \right )
        }{ \mathfrak{a}_k^{(l)}[n]
        } \right )\\
    &-\sum_{j \in \{\mathcal{K}[n] \backslash k\}}
    \mathrm{tr} 
     \left ( \frac{
            \log_2(\mathrm{e}) \left (
            \sigma^2 \left ( \boldsymbol{W}_k[n]-\boldsymbol{W}_k^{(l)}[n] \right )
            \right )
        }{ \mathfrak{a}_k^{(l)}[n]
        } \right ) \triangleq \check{R}_k [n] (\boldsymbol{W}_k)
    ,
    \end{split}
    \label{sca}}
\end{equation}
where \scalebox{0.84}{$\mathfrak{a}_k^{(l)}[n] = \sum_{j \in \{\mathcal{K} \backslash k\}} \mathrm{tr} \left ( P_j[n]\boldsymbol{H}_j[n] \boldsymbol{W}_k^{(l)}[n] \right  ) + \mathrm{tr}\left ( \sigma^2\boldsymbol{W}_k^{(l)}[n] \right  )$}. Therefore, the objective function of (P3) and the constraint \eqref{p3c} can be replaced by its lower bound as
{\begin{subequations} {
\begin{flalign}
(\text{P3}l)\ \max_{\boldsymbol{W}} \quad & \sum_{k\in \mathcal{K}[n]} \check{R}_k [n] (\boldsymbol{W}_k) \nonumber\\
 {\rm{s.t.}}  \quad
&\text{(\ref{p3a}) and (\ref{p3b})}, \nonumber\\
&\sum_{n\in \Bar{\mathcal{N}_k}} \check{R}_k[n](\boldsymbol{w}_k, \boldsymbol{q}, P_k) \geq \frac{D_{\mathrm{th}}}{\tau B }, \quad \forall k \in
\mathcal{K}.\label{p4b}
\end{flalign}}\end{subequations}}Note that the constraint (\ref{p3b}) is a non-convex rank constraint, which can be relaxed via semi-definite relaxation (SDR), and the relaxed problem (P3$l$.SDR) is denoted as
{\begin{subequations} {
\begin{flalign}
(\text{P3}l\text{.SDR})\ \max_{\boldsymbol{W}} \quad & \sum_{k\in \mathcal{K}[n]} \check{R}_k [n] (\boldsymbol{W}_k) \nonumber\\
 {\rm{s.t.}}  \quad
&\text{(\ref{p3a}) and (\ref{p4b})}, \nonumber\\
&\boldsymbol{W}_k[n] \succeq \boldsymbol{0}, \quad \forall k\in \mathcal{K}[n].
\end{flalign}}\end{subequations}}The problem (P3$l$.SDR) is a standard convex optimization problem.
Let {$\{ \boldsymbol{W}_k^{\star}, k \in \mathcal{K}[n] \}$} denote the optimal solution.
If the rank of the solution satisfies {$\mathrm{rank} (\boldsymbol{W}_k^{\star}) \leq 1$}, {$ \boldsymbol{W}_k^{\star}$} is also a solution to (P3$l$). Otherwise, 
we introduce the following proposition to prove that the solution to (P3$l$.SDR) is also a solution to (P3$l$).

\begin{proposition}
There always exists a rank-one optimal solution to (P3$l$.SDR), which is denoted as \scalebox{0.94}{$\{\boldsymbol{\bar{W}}_k^{}, \forall k \in \mathcal{K}[n] \}$}.
\begin{proof}
\renewcommand{\qedsymbol}{}
Let \scalebox{0.94}{$\bar{\boldsymbol{w}}_k^{} = \left( 
    \boldsymbol{a}_k^{\mathsf{H}} \boldsymbol{W}_k^{\star}\boldsymbol{a}_k
    \right)^{-\frac{1}{2}} \boldsymbol{W}_k^{\star}\boldsymbol{a}_k$}. Then we have \scalebox{0.94}{$\boldsymbol{\bar{W}}_k = (\bar{\boldsymbol{w}}_k^{} )^{\mathsf{H}}\bar{\boldsymbol{w}}_k^{}$}. It is obvious that \scalebox{0.94}{$\boldsymbol{\bar{W}}_k^{}$} is positive semi-definite, and is rank-one. By substituting \scalebox{0.94}{$\boldsymbol{\bar{W}}_k^{}$} into \eqref{sca}, we can derive that \scalebox{0.94}{$\check{R}_k [n] (\boldsymbol{\bar{W}}_k) = \check{R}_k [n] (\boldsymbol{W}^{\star}_k)$}. Hence, \scalebox{0.94}{$\boldsymbol{\bar{W}}_k^{}$} is verified to be the optimal solution to (P3$l$.SDR). Since \scalebox{0.94}{$\boldsymbol{\bar{W}}_k^{}$} is rank-one, it is also the optimal solution to (P3$l$).
\end{proof}
\end{proposition}

\subsection{Overall Algorithm and Performance Analysis}
The proposed double-loop RLA is illustrated in Algorithm \ref{rla}, where $E_{\max}$ is the maximum number of training episodes of the RLA.
Since the flying region is fixed and the number of GNs is determined, the reward function is bounded, which ensures the convergence of the outer-loop FDQN algorithm. Besides, the objective of problem (P3$l$.SDR) is non-decreasing, thus the convergence of the inner-loop SCA is ensured. In summary, the convergence of the proposed RLA can be guaranteed.

{The computational complexity of the proposed algorithm is analyzed as follows. Firstly, in the outer loop of the DRL algorithm, for each training batch, 2 + $K$ is the action space dimension, and $ ((2 + K) G) ^ 2 $ is derived from the Gaussian noise parameter update of each action dimension in the noise network. The network has two hidden layers, and the numbers of nerve cells in the two hidden layers are $g_1$ and $g_2$. The number of nerve cells in the hidden layer determines the number of matrix multiplications. The complexity is $g_1 ^ 2 + g_2 ^ 2 $. $\Xi $ and $\log\zeta $ correspond to the overhead of partial learning and priority sampling, respectively. $U$ corresponds to the number of training episodes. So the computational complexity of the entire training process is $\mathcal{O}\left(UN ((2 + K) G) ^ {2} + g_{1} ^ {2} + g_{2}^{2} + G^{2} +\Xi +\log\zeta \right)$. Next, we will analyze the complexity of the inner-loop. The computational complexity for solving a convex optimization problem by the inner point method is $\mathcal{O}\left ((m) ^ {3.5}\log\frac {1} {\epsilon}\right) $, where $m$ is the dimension of the optimization variable. Based on this result, since the dimension of the optimization variable is $K\times N$, the computational complexity for solving problem P3l.SDR is $\mathcal{O}\left ((KN) ^ {3.5}\log\frac {1} {\epsilon}\right) $. Therefore, the computational complexity of the RLA algorithm is
$ \mathcal{O}  ( UN ( \mathcal{L}
(KN) ^ {3.5}\log\frac {1} {\epsilon} + ((2 + K) G) ^ {2} + g_ {1} ^ {2} + g_ {2} ^ {2} + G ^ {2} $
$ + \Xi +\log\zeta) ) $, where $\mathcal{L}$ is the number of iterations of the SCA.}

\begin{algorithm} \scriptsize
	\caption{RLA for Solving (P1)}
	\label{rla}
	\begin{algorithmic}
\REQUIRE {The position of the UAV $\boldsymbol{q}[1]$, the position of GNs $\boldsymbol{s}_k, \forall k$;}\\
\textbf{Initialize:} the evaluate network $\theta$, the target network $\theta'$, the PER;\\
		
\FOR{Episode = $\{1,2,\cdots,E_{\max} \}$}
\WHILE{$n\neq N$}
		\STATE Observe the state ${s}[n] = (\boldsymbol{q}[n], \boldsymbol{H}[n], \boldsymbol{s}_{k})$;\\
            \STATE Choose the action ${a}[n] = (\Delta_{\text{UAV}}[n], \theta_{\text{UAV}}[n],  \boldsymbol{P}_{\text{}}[n])$ in the outer-loop;\\
            \STATE Solve the problem (P2) in the inner-loop;\\
            \STATE {Calculate the reward value $r[n]$ by \eqref{calculate_reward};}\\
            \STATE Observe the next state ${s}[n+1]$;\\
            \STATE Store transition $({s}[n], {a}[n], r[n], {s}[n+1])$ in the PER;\\
            \STATE Sample random minibatch of transitions from PER;\\
            \STATE Update the evaluate network $\theta$ through target value $y^{'}$;\\
            \STATE Update the target network by ${\theta}^{'} \leftarrow {\theta}$ according to the update rate $\Bar{\tau}$;\\
\ENDWHILE\\
Reset the position of the UAV;
\ENDFOR
\ENSURE {The trained networks $\theta^*$ and $\theta^{\prime*}$.}\\
	\end{algorithmic}
\end{algorithm}

\subsection{All Learned Once by FDQN}
To obtain more insights of the proposed FDQN structure, we introduce a
FDQN-based algorithm called all learned once (ALO),
where the receive beamforming of the UAV $\{\boldsymbol{w}_k [n], \forall k \in \mathcal{K}[n] \}$ is also considered as an action within the FDQN structure.

In the proposed ALO, we define the MDP as follows: the observation state space $\mathcal{S}^{\prime}[n], \forall n$, the action space $\mathcal{A}^{\prime}[n], \forall n$, and the reward function $R^{\prime}[n], \forall n$.
Similar to \cite{9417469, 10433196}, we introduce a codebook $\mathcal{W}$ to design the actions for optimizing the receive beamforming of the UAV. Hence, the action space of the ALO is rewritten as
\begin{equation} 
\mathcal{A}^{\prime}[n] = (\Delta_{\text{UAV}}[n], \theta_{\text{UAV}}[n], \{\boldsymbol{w}_k [n], \forall k \in \mathcal{K}[n] \}, \boldsymbol{P}_{\text{}}[n]),\ \forall n,
\end{equation}
where \scalebox{0.94}{$\{\boldsymbol{w}_k [n], \forall k \in \mathcal{K}[n] \}$} is chosen from the codebook \scalebox{0.94}{$\mathcal{W}$}.
Furthermore, the observation state space $\mathcal{S}^{\prime}$ and the reward function $R^{\prime}$ are identical to those in the RLA, i.e., \scalebox{0.94}{$\mathcal{S}^{\prime}[n] \triangleq \mathcal{S}[n], \forall n$} and \scalebox{0.94}{$R^{\prime}[n] \triangleq R[n], \forall n$}.

{{The ALO algorithm does not require an inner SCA, but adds the selection of the beamforming matrix to the action space, so the action space in the ALO algorithm is 2 + 2$K$, so the computational complexity of the ALO algorithm is $\mathcal{O}\left (UN ((2 + 2K) G) ^ {2} + g_ {1} ^ {2} + g_ {2} ^ {2} + G ^ {2} +\Xi +\ log\zeta\right) $.}}

\begin{algorithm} \scriptsize
	\caption{ALO for Solving (P1)}
	\label{alo}
	\begin{algorithmic}
\REQUIRE {The position of the UAV $\boldsymbol{q}[1]$, the position of GNs $\boldsymbol{s}_k, \forall k$;}\\
\textbf{Initialize:} the evaluate network $\theta$, the target network $\theta'$, the PER;\\
		
\FOR{Episode = $\{1,2,\cdots,E_{\max} \}$}
\WHILE{$n\neq N$}
		\STATE Observe the state ${s}[n] = (\boldsymbol{q}[n], \boldsymbol{H}[n], \boldsymbol{s}_{k})$;\\
            \STATE Choose the action ${a}[n] = (\Delta_{\text{UAV}}[n], \theta_{\text{UAV}}[n], \{\boldsymbol{w}_k [n]\},  \boldsymbol{P}_{\text{}}[n])$;\\
            \STATE {Calculate the reward value $r[n]$ by \eqref{calculate_reward};}\\
            \STATE Observe the next state ${s}[n+1]$;\\
            \STATE Store transition $({s}[n], {a}[n], r[n], {s}[n+1])$ in the PER;\\
            \STATE Sample random minibatch of transitions from PER;\\
            \STATE Update the evaluate network $\theta$ through target value $y^{'}$;\\
            \STATE Update the target network by ${\theta}^{'} \leftarrow {\theta}$ according to the update rate $\Bar{\tau}$;\\
\ENDWHILE\\
Reset the position of the UAV;
\ENDFOR
\ENSURE {The trained networks $\theta^*$ and $\theta^{\prime*}$.}\\
	\end{algorithmic}
\end{algorithm}

\section{Numerical Results}
\begin{figure}[!h]
	\begin{minipage}{0.475\linewidth}
		\centering
		\includegraphics[width=1\linewidth]{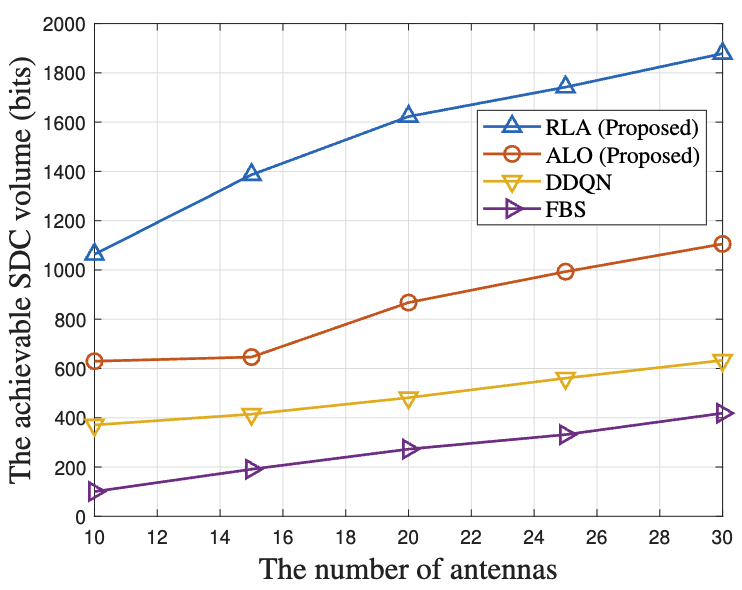}
		\captionsetup{font=small}
            \caption{\centering{The SDC versus the antennas number of the UAV.
            }}
		\label{antenna}
	\end{minipage}
	\begin{minipage}{0.475\linewidth}
		\centering
		\includegraphics[width=1\linewidth]{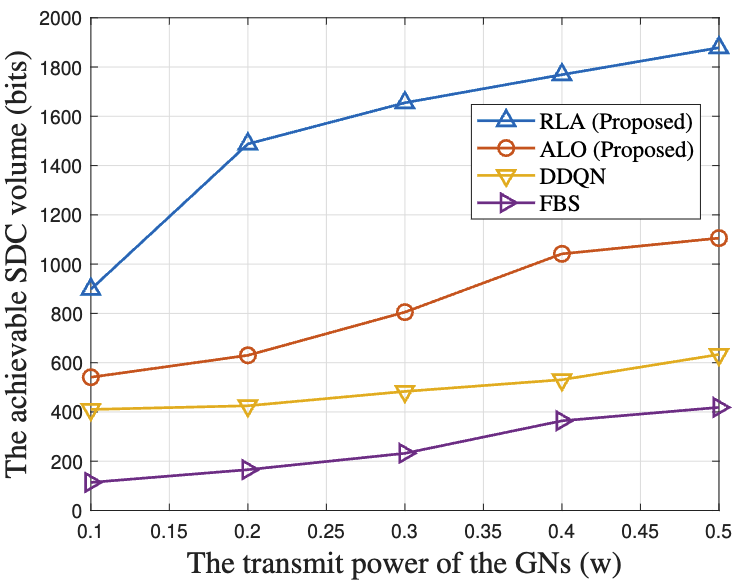}
            \captionsetup{font=small}
		\caption{\centering{The SDC versus the transmit power threshold of the GNs.
            }}
		\label{power}
	\end{minipage}
\end{figure}

\begin{figure}[!h]
	\begin{minipage}{0.475\linewidth}
		\centering
		\includegraphics[width=1\linewidth]{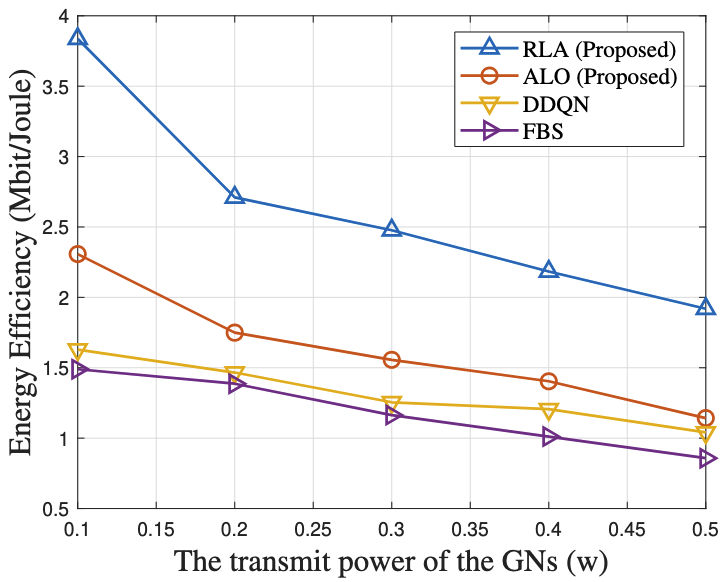}
		\captionsetup{font=small}
            \caption{\centering{ The Energy Efficiency versus the transmit power of the GNs.
            }}
		\label{energy efficiency}
	\end{minipage}
	\begin{minipage}{0.475\linewidth}
		\centering
		\includegraphics[width=1\linewidth]{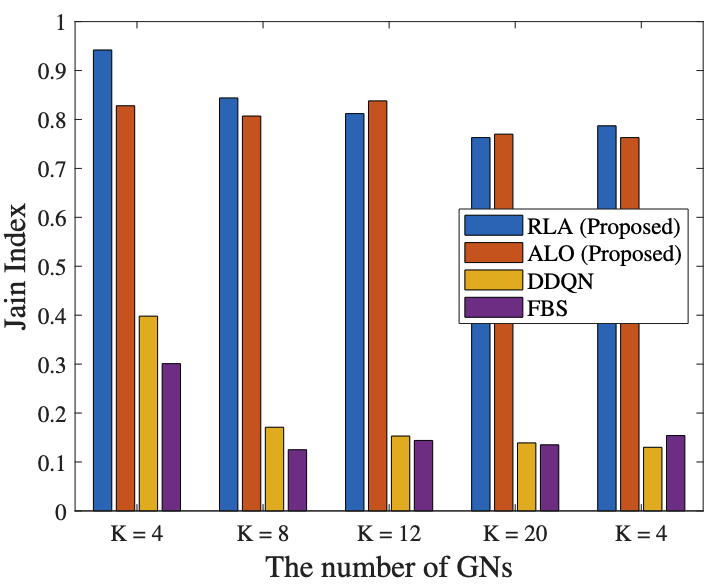}
            \captionsetup{font=small}
		\caption{\centering{The Jain Index versus the number of the GNs.
            }}
		\label{fairness_index}
	\end{minipage}
\end{figure}

This section compares the numerical results of the proposed algorithms with two benchmarks, i.e., {the ALO algorithm based on double DQN (DDQN)} and fixed beamforming selection (FBS).
The flying region is a space with \((x=1,000 \mathrm{m}, y=1,000 \mathrm{m})\). The main parameters are summarized as follows: the number of GNs \(K\) is $20$, the number of antennas of the UAV \(M\) is $30$, the maximum and minimum flight speed of the UAV \(V_{\max}\) and $V_{\min}$ is $17$m/s and $5$m/s, respectively, the flying height of the UAV is $50$m,
the number of time slots \(N\) is $30$,
and the bandwidth $B$ is $80$MHz. The hyper-parameters of the RLA and the ALO are as follows: the learning rate is $0.0005$, the reward decay rate is $0.9$, the target network update frequency is $10$ with update rate $\Bar{\tau} = 0.05$. ReLU and Adam are utilized as the activation function and optimizer, respectively.
{Meanwhile, the numerical results in Figs.~\ref{antenna}, Figs.~\ref{power}, Figs.~\ref{energy efficiency}, and Figs.~\ref{fairness_index} are the average values obtained from $10$ simulations.}
\begin{figure}[!h]
	\begin{minipage}{0.475\linewidth}
		\centering
		\includegraphics[width=1\linewidth]{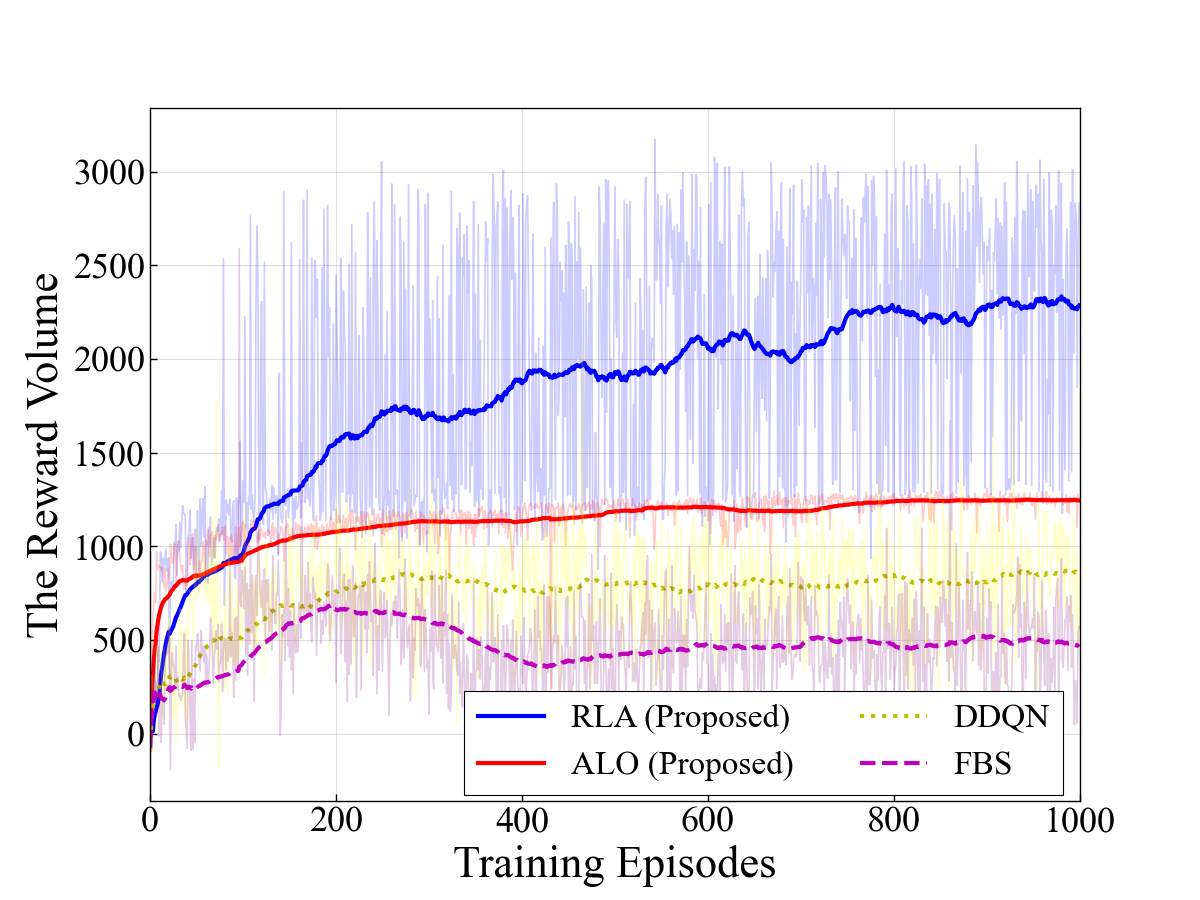}
		\captionsetup{font=small}
            \caption{\centering{The reward value versus the training episodes.
            }}
		\label{reward}
	\end{minipage}
	\begin{minipage}{0.475\linewidth}
		\centering
		\includegraphics[width=1\linewidth]{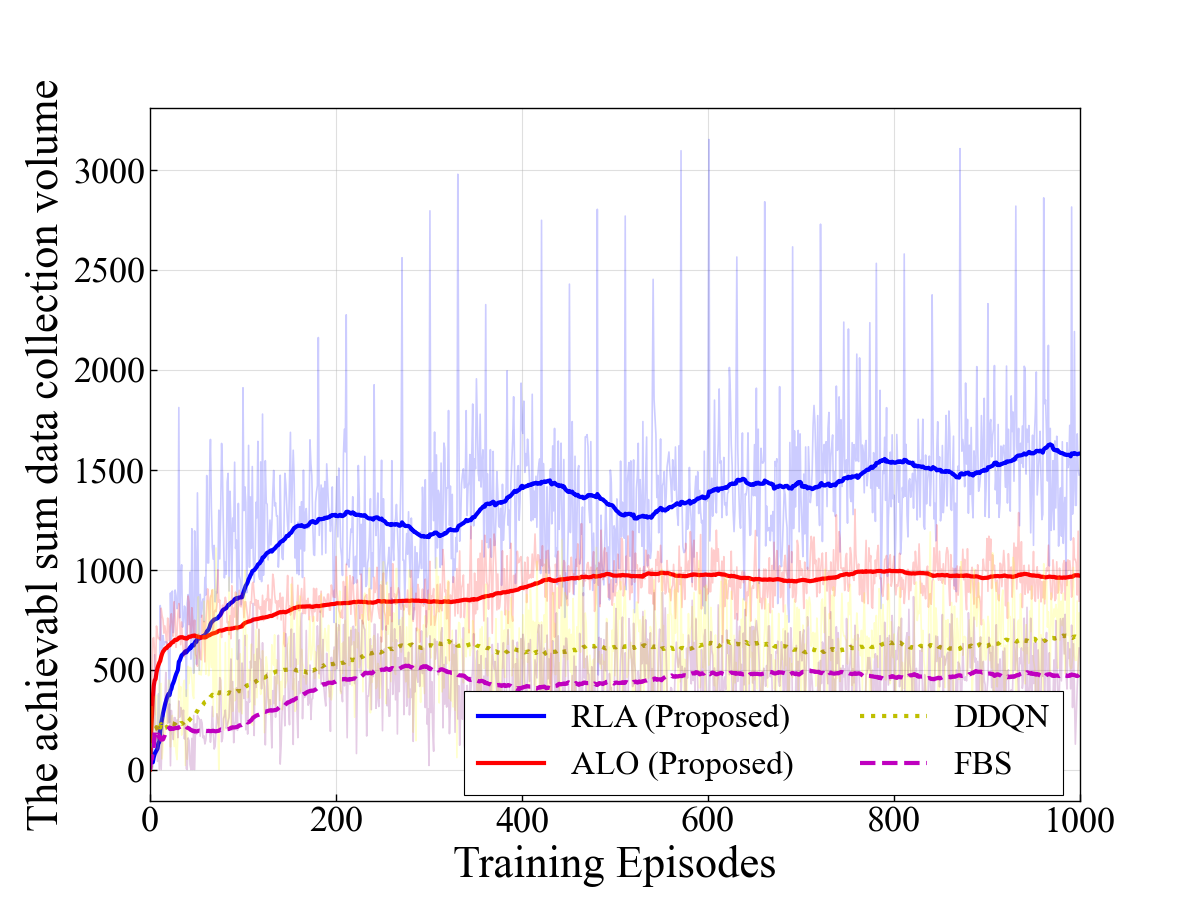}
            \captionsetup{font=small}
		\caption{\centering{The SDC volume versus the training episodes.
            }}
		\label{data}
	\end{minipage}
\end{figure}

Fig.~\ref{reward} and Fig.~\ref{data} illustrate
the reward value versus the training episodes, and the achievable SDC volume versus the training episodes, respectively.
The RLA achieves the best performance in terms of reward value. The reason is that after training, the trajectory of the UAV is optimized in conjunction with decisions on the GN's scheduling and resource allocation.
Additionally, the SCA-enabled RLA provides the optimal receive beamforming of the UAV, enabling larger reward than other methods.
Correspondingly, the achievable SDC of the RLA is the highest among all methods.
Furthermore, the achieved reward of the ALO is similar to that of the RLA.
The reason is that, the DRL algorithm adapts to select actions that optimize the reward value based on the current system state.
However, in terms of the SDC performance, the ALO is less effective than the RLA.
This is due to the discrete beamforming codebook selection in the ALO, which may overlook superior beamforming actions, whereas the RLA can provide optimal beamforming optimization.

Fig.~\ref{antenna} and Fig.~\ref{power} illustrate the achievable SDC volume versus the number of antennas of the UAV, and the transmit power of the GNs, respectively.
In Fig.~\ref{antenna}, as the number of antennas increases, the channel gains between the GNs and the UAV improve, leading to enhanced achievable SDC for all methods. The curves of different algorithms have similar trend as Fig.~\ref{power} shows. Moreover, our proposed RLA demonstrates superior performance, benefiting from the optimal receive beamforming optimized in the inner-loop, and the UAV trajectory, scheduling, and resource allocation optimized in the {outer-loop}.
Additionally, the proposed ALO achieves performance close to the RLA, demonstrating the adaptability and deployment prospects of the FDQN-based ALO algorithm {with lower complexity compared with the RLA.}

{Fig.~\ref{energy efficiency} illustrates the relationship between the energy efficiency and the transmit power of the GNs. It can be seen from Fig.~\ref{energy efficiency} that the proposed RLA algorithm and the ALO algorithm are significantly superior to the comparative algorithms in terms of energy efficiency. As the transmit power increases, the interference in the simultaneous communications between the UAV and multiple GNs increases, and the overall energy efficiency shows a downward trend. Fig.~\ref{fairness_index} demonstrates the fairness indicators among the scheduling of various GNs under different numbers of GNs. We utilize the widely adopted Jain Index to measure fairness, which can be expressed as \cite{jain1984quantitative}:
\begin{equation}
    J\left({\mathcal{D}}_{1}, {\mathcal{D}}_{2}, \ldots, {\mathcal{D}}_{K}\right)=\frac{\left(\sum_{k=1}^{K} {\mathcal{D}}_{k}\right)^{2}}{K \cdot \sum_{k=1}^{K}\left({\mathcal{D}}_{k}\right)^{2}}
\end{equation}
where ${\mathcal{D}}_k$ represents the total collected data volume from the $k$-th GN, i.e., ${\mathcal{D}}_k = \sum_{n=1}^N D_k[n](\boldsymbol{w}_k, \boldsymbol{q}, P_k) $.
We can observe from Fig.~\ref{fairness_index} that under different numbers of the GNs, the proposed RLA algorithm and the ALO algorithm exhibit excellent performance in the fairness of GNs.}

\begin{figure}[!h]
	\begin{minipage}{0.485\linewidth}
		\centering
		\includegraphics[width=1\linewidth]{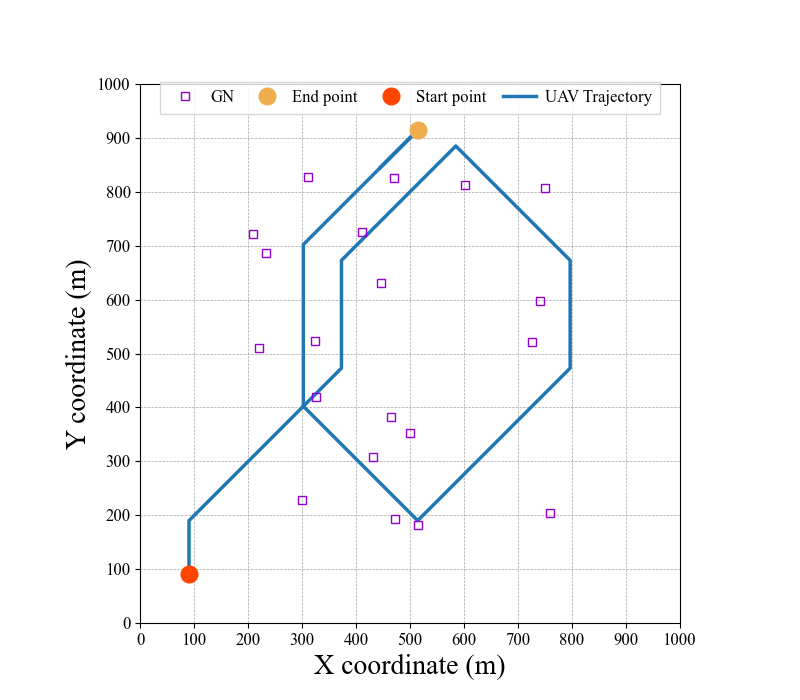}
		\captionsetup{font=small}
            \caption{\centering{The trajectory of the UAV from the 2D perspective.
            }}
		\label{2d}
	\end{minipage}
	\begin{minipage}{0.485\linewidth}
		\centering
		\includegraphics[width=1\linewidth]{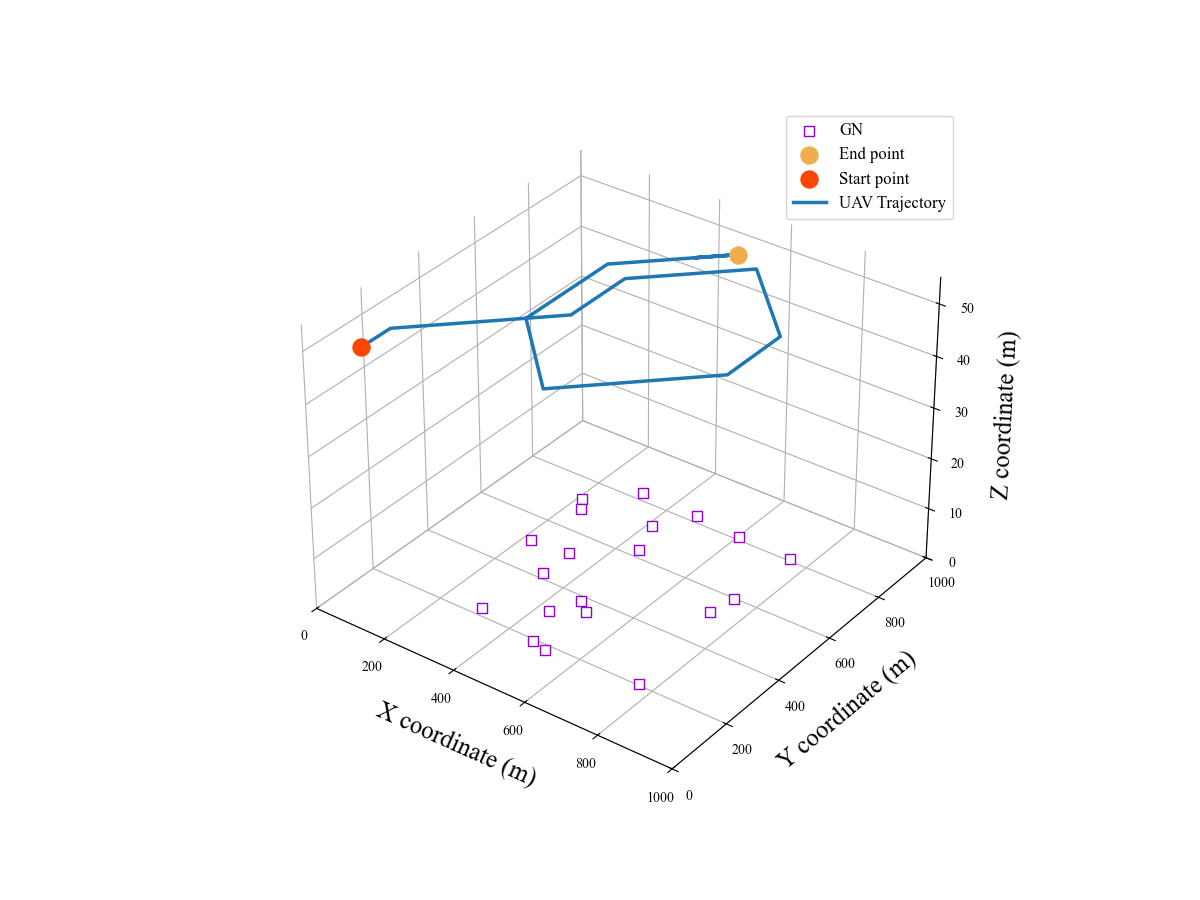}
            \captionsetup{font=small}
		\caption{\centering{The trajectory of the UAV from the 3D perspective.
            }}
		\label{3d}
	\end{minipage}
\end{figure}

The UAV trajectories in both 2D and 3D views of the proposed RLA are shown in Fig.~\ref{2d} and Fig.~\ref{3d}, respectively.
We can observe that the RLA agent can select a reasonable trajectory based on the locations of GNs, ensuring the maximization of the SDC.

\section{Conclusion}
In this paper, we considered a UAV-enabled data collection system
to maximize the achievable SDC of all GNs.
The trajectory of the UAV, the receive beamforming of the UAV, the scheduling and the power allocation of all GNs are jointly optimized.  
Two DRL-based algorithms are proposed, where the first proposed RLA algorithm leverages the strengths of optimization and the DRL, while the second proposed ALO algorithm optimizes the variables throughout the DRL.
Numerical results verify the effectiveness of the proposed algorithms comparing with two benchmark algorithms. Due to the optimal beamforming obtained by the optimization method, the RLA algorithm always achieves a much better performance than the ALO algorithm in terms of the SDC. {Additionally,
multi-UAV systems are studied for extended coverage.
However, their implementation introduces significant complexities in coordinated control and resource allocation.
Therefore, future works could explore more efficient beamforming optimization algorithms to reduce computational complexity, and combine federated learning or transfer learning to reduce dependence on large-scale training data, and further improve applicability.}

\bibliographystyle{IEEEtran}
\bibliography{ref}
\end{document}